\documentclass[12pt,preprint]{aastex}

\def\gs{\mathrel{\raise0.35ex\hbox{$\scriptstyle >$}\kern-0.6em \lower0.40ex\hbox{{$\scriptstyle \sim$}}}}
\def\ls{\mathrel{\raise0.35ex\hbox{$\scriptstyle <$}\kern-0.6em \lower0.40ex\hbox{{$\scriptstyle \sim$}}}}
\newcommand{\Msolar}{\mbox{$M_{\odot}\,$}}
\newcommand{\Lsolar}{\mbox{$L_{\odot}\,$}}

\newcommand{\arcsecs}{\mbox{$^{\prime\prime}$}}

\shorttitle{Dense Gas in Luminous Submillimeter Galaxies}
\shortauthors{Greve et al.}

\begin{document}


\title{A Search for Dense Gas in Luminous Submillimeter Galaxies with the 100-m Green Bank Telescope}


\author{T.\ R.\ Greve\altaffilmark{1}}
\email{tgreve@submm.caltech.edu}
\author{L.\ J.\ Hainline\altaffilmark{1}}
\author{A.\ W.\ Blain\altaffilmark{1}}
\affil{California Institute of Technology, Pasadena, CA 91125, USA}
\author{Ian Smail\altaffilmark{2}}
\affil{Institute for Computational Cosmology, University of Durham, South Road, Durham DH1 3LE, United Kingdom}
\author{R.\ J.\ Ivison \altaffilmark{3}}
\affil{UK ATC, Royal Observatory, Blackford Hill, Edinburgh EH9 3HJ, United Kingdom}
\author{P.\ P.\ Papadopoulos \altaffilmark{4}}
\affil{Institut f\"{u}r Astronomie, ETH, Z\"{u}rich, 8093 Switzerland}






\begin{abstract}
We report deep K-band (18-27\,GHz) observations with the 100-m Green Bank Telescope 
of HCN$(1-0)$ line emission towards the two submillimeter-selected galaxies (SMGs)
SMM\,J02399$-$0136 ($z=2.81$) and SMM\,J16359$+$6612 ($z=2.52$).
For both sources we have obtained spectra with channel-to-channel rms noise of $\sigma \le 0.5$\,mJy,
resulting in velocity-integrated line fluxes better than $\ls 0.1$\,Jy\,km\,s$^{-1}$, although we
do not detect either source.
Such sensitive observations -- aided by gravitational lensing of the sources -- permit us
to put upper limits of $L'_{\mbox{\tiny{HCN}}}\ls 2\times 10^{10}\,$K\,km\,s$^{-1}\,$pc$^2$ 
on the intrinsic HCN$(1-0)$ line luminosities of the two SMGs.
The far-infrared (FIR) luminosities for all three SMGs with sensitive HCN$(1-0)$ observations
to date are found to be consistent with the tight FIR-HCN luminosity correlation observed
in Galactic molecular clouds, quiescent spirals and (ultra) luminous infrared galaxies in the
local Universe. Thus, the observed HCN luminosities remain in accordance with what
is expected from the universal star formation efficiency per {\it dense} molecular gas
mass implied by the aforementioned correlation, and more sensitive observations
with today's large aperture radio telescopes hold the promise
of detecting HCN$(1-0)$ emission in similar objects in the distant Universe.
\end{abstract}





\section{Introduction}
The search for molecular lines at high redshifts ($z\gs 1$) offers
one of the richest and most exciting avenues to follow in the study
of galaxy formation and evolution (Solomon \& Vanden Bout 2005).
The detections of rotational transitions of CO in extremely distant quasars 
(Carilli et al.\ 2002; Walter et al.\ 2003), high-z radio galaxies 
(Papadopoulos et al.\ 2000; De Breuck et al.\ 2005) and submillimeter (submm) 
selected galaxies (Frayer et al.\ 1998, 1999; Neri et al.\ 2003), have 
established that the most luminous and extreme objects in the early Universe 
harbour vast amounts of molecular gas ($\sim 10^{10-11}\,\Msolar$ -- Neri et al.\ 2003; Greve et al.\ 2005) 
and have large dynamical masses ($\sim 10^{11}\,\Msolar$ -- Genzel et al.\ 2003; Tacconi et al.\ 2006). 
Modelling of the bulk physical conditions of the molecular gas in high-$z$ QSOs
have even been possible in cases where several CO lines have been detected
(Wei\ss\ et al.\ 2005). The success of CO observations is primarily due to its much larger
abundance compared to other species as well as its easily excitable low-$J$ 
(CO $J+1\longrightarrow J, J<3$) rotational lines, which makes it
an excellent tracer of the total amount of {\it diffuse}
($n(\mbox{H}_2) \sim 10^{2-3}$\,cm$^{-3}$) molecular gas.
However, the diffuse gas only serves as a reservoir of {\it potential} fuel
available for star formation, and is not actively involved in forming stars.

In our own Galaxy, the sites of active star formation coincide with the 
dense cores of giant molecular clouds (GMCs). This dense gas
phase ($n(\mbox{H}_2) \gs 10^{4}$\,cm$^{-3}$) is best traced
using molecules with high critical densities such as HCN.
This basic picture also appears to hold in nearby galaxies, 
as suggested by HCN$(1-0)$ surveys of local ($z\ls 0.1$) Luminous Infrared Galaxies
(LIRGs -- $L_{\mbox{\tiny{IR}}}\sim 10^{11}\,\Lsolar$) and Ultra
Luminous Infrared Galaxies (ULIRGs -- $L_{\mbox{\tiny{IR}}}\sim 10^{12}\,\Lsolar$),
which found a remarkably tight correlation between IR luminosity
and HCN line luminosity (Solomon et al.\ 1992; Gao \& Solomon 2004a,b).
Recently, this relation was shown to extend all the way down to
single GMCs, thus extending over 7-8 orders of magnitude in IR luminosity 
(Wu et al.\ 2005). This was interpreted as suggesting that the true star formation efficiency
(i.e.\ the star formation rate per unit mass of {\it dense} gas) is the same 
in these widely different systems. 

Attempts have been made at extending the correlation between IR and HCN luminosity to primordial
galaxies at high redshifts 
(Barvainis et al.\ 1997; Solomon et al.\ 2003; Isaak et al.\ 2004; Carilli et al.\ 2005), 
resulting in four objects detected in HCN out of 8 objects observed.
So far the results have been consistent with the local correlation,
however, the small number of sources precludes a statistically
robust conclusion as to whether the relation also holds at high redshift. 
Furthermore, the majority of high-$z$ objects observed have
been extremely luminous QSOs and HzRGs dominated by Active Galactic Nuclei (AGN).

The detection of HCN$(1-0)$ in the $z=2.286$ QSO F\,10214$+$4724 
(Vanden Bout, Solomon \& Maddalena 2004), which marked the first 
high-$z$ detection of HCN using the Green Bank Telescope (GBT),
motivated us to test the capability of the GBT to probe the dense 
gas in a population of high-$z$ galaxies less
extreme, i.e.\ less AGN-dominated than F\,10214$+$4724, 
but more representative of starburst galaxies in the early Universe.
The goal of our study was to observe two of the most prominent examples of
submillimeter-selected galaxies (see Blain et al.\ 2002 for 
a review) -- a population likely to constitute the progenitors of today's 
massive spheroids, and thus fundamental to our understanding of galaxy formation and evolution.

Throughout this paper we adopt a flat cosmology, with $\Omega_m=0.27$, $\Omega_\Lambda=0.73$ and
$H_0=71$\,km\,${\mbox{s}^{-1}}$\,Mpc$^{-1}$ (Spergel et al.\ 2003).

\section{Sources}\label{section:sources}
The two submillimeter galaxies targeted with the GBT were
SMM\,J02399$-$0136 ($z=2.808$) and SMM\,J16359$+$6612 ($z=2.517$) 
\footnote{hereafter we shall refer to these three sources
as J02399 and J16359, respectively}.
These two sources were chosen by virtue of
1) their gravitational lensing amplifications making them
amongst the brightest submm sources known
(Smail et al.\ 1997, 2002; Kneib et al.\ 2004), and 2)
their strong CO line emission
(Frayer et al.\ 1998; Genzel et al.\ 2003; Sheth et al.\ 2004; Kneib et al.\ 2005).
Moreover, while J02399 has a moderate amplification
factor ($\mu \sim 2.5$) and thus is intrinsically very luminous, 
the extremely large amplification factor of 
J16359 ($\mu \sim 22$) means that this source has
about an order of magnitude smaller intrinsic luminosity.
Thus, our observations could potentially explore differences
in the dense gas mass fraction between massive and not so massive
starburst galaxies at high redshifts.

\section{Observations and Data Reduction}\label{section:observations-and-data-reduction}
The observations were carried out between 
March 21st and April 5th 2005 at
the 100-m Robert C.\ Byrd Green Bank Telescope 
in Green Bank, West Virginia\footnote{The GBT is
operated by the National Radio Astronomy Observatory. The National Radio
Astronomy Observatory is a facility of the National Science Foundation operated under
cooperative agreement by Associated Universities, Inc.}.
Due to scheduling and observability constraints, J02399 received by far 
the most observing time: 9.6\,hrs of effective integration in each
polarisation, with J16359 getting only 3.3\,hrs.

At the redshifts of the two sources, the HCN$(1-0)$ line 
($\nu_{\mbox{\tiny{rest}}} = 88.631$\,GHz) falls within the 
K-band (17.6-27.1\,GHz) receiver on the GBT. The K-band receiver 
consists of two pairs of dual circular polarisation beams,
covering the frequency ranges 17.6-22.4\,GHz and 22.0-27.1\,GHz, respectively.
In our case, the HCN$(1-0)$ line from both sources could be observed
using only one pair of beams (22.0-27.1\,GHz).
At $\sim 25$\,GHz the GBT has a beam size of {\sc fwhm}~$=30\arcsecs$ -- 
much larger than the source size ($\ls 2 \arcsecs$).

The observations were made in dual-beam mode, in which one
feed sees the source ({\tt ON}) and the other sees the blank sky ({\tt OFF}).
After one such scan (which in our case lasted 2 minutes),
the telescope nods so that the first feed now is {\tt OFF} and the second
feed is {\tt ON} the source, thus ensuring that
one beam is always integrating on-source (except during the nodding).
As backend we used the GBT Spectrometer in its moderate-bandwidth, low-resolution
submode, giving 200\,MHz of total instantaneous bandwidth divided into
8192 channels each of width 24.41\,kHz. At 25\,GHz this corresponds to
a velocity-coverage of $\simeq 2400$\,km\,s$^{-1}$ and a resolution
of $\simeq 0.3$\,km\,s$^{-1}$ per channel -- more than enough to encompass
the entire line and resolve it spectrally. 
This is based on the reasonable assumption that the HCN line widths are similar to
those of CO (Greve et al.\ 2005).

Calibration was by noise-injection, in which a noise-diode is
switched on and off at a rate of 1\,Hz during the integration. 
Thus a typical data time stream can be visualised as follows:\\

{\small
\hspace*{4cm}\hspace*{1.88cm} {\tt Beam 1} \hspace*{2.08cm} {\tt Beam 2}\\
\\
\hspace*{4cm}\hspace*{2.65cm}${\tt ON}_{1,1}$\hspace*{2.55cm}${\tt OFF}_{2,1}$\\
\hspace*{4cm}{\tt Scan 1:}\hspace*{0.45cm}        $\overbrace{[\mbox{{\tt ON}}_{calon},\mbox{{\tt ON}}_{caloff}]}$ \hspace*{0.5cm} $\overbrace{[\mbox{{\tt OFF}}_{calon},\mbox{{\tt OFF}}_{caloff}]}$\\ 
\hspace*{4cm}\hspace*{2.48cm} ${\tt OFF}_{1,2}$ \hspace*{2.28cm} ${\tt ON}_{2,2}$\\
\hspace*{4cm}{\tt Scan 2:}\hspace*{0.257cm}        $\overbrace{[\mbox{{\tt OFF}}_{calon},\mbox{{\tt OFF}}_{caloff}]}$ \hspace*{0.63cm}$\overbrace{[\mbox{{\tt ON}}_{calon},\mbox{\tt ON}_{caloff}]}$\\ 
\hspace*{4cm}\hspace*{2.65cm}${\tt ON}_{1,3}$\hspace*{2.55cm}${\tt OFF}_{2,3}$\\
\hspace*{4cm}{\tt Scan 3:}\hspace*{0.45cm}        $\overbrace{[\mbox{{\tt ON}}_{calon},\mbox{{\tt ON}}_{caloff}]}$ \hspace*{0.5cm} $\overbrace{[\mbox{{\tt OFF}}_{calon},\mbox{{\tt OFF}}_{caloff}]}$\\ 
\hspace*{4cm}.\\
\hspace*{4cm}.\\
\hspace*{4cm}.\\
}

The system temperature on the blank sky is then given by:
\begin{equation}
T_{sys}(\nu,p) = T_{cal}(\nu,p) \frac{\mbox{{\tt OFF}}_{caloff}}
{\mbox{{\tt OFF}}_{calon}-\mbox{{\tt OFF}}_{caloff}} + \frac{T_{cal}(\nu,p)}{2},
\label{equation:calibration}
\end{equation}
where $T_{cal}(\nu,p)$ is the calibration temperature of the noise diode and
is a function of frequency ($\nu$) and polarisation ($p$). Instead of $T_{cal}(\nu,p)$
we adopted the mean $T_{cal}(p)$ value across the entire bandpass.
Throughout the observing run the system temperature (integrated over the entire
bandwidth observed) varied between
28\,K in optimal conditions to 55\,K in the worst cases, depending on elevation and
the water vapour content of the atmosphere. 
Each scan was inspected by eye in order to weed out
bad or abnormal looking scans.

The data were reduced using two independent methods. The first method 
used the GBTIDL-routine GETNOD, which takes the {\tt OFF} signal from one scan and subtracts it from
the {\tt ON} signal in the neighbouring scan, and vice versa.
In this scheme the spectra are calibrated onto the antenna temperature scale following:
\begin{equation}
T_{ant,j}(\nu,p) = T_{sys}(\nu,p) \frac{{\tt ON_{i,j}}-{\tt OFF_{i,j+1}}}{{\tt OFF_{i,j+1}}},  \\
\end{equation}
where ${\tt ON_{i,j}} = ({\tt ON}_{calon}+{\tt ON}_{caloff})/2$ and
${\tt OFF_{i,j+1}} = ({\tt OFF}_{calon}+{\tt OFF}_{caloff})/2$, and $i$, $j$
represent the beam and scan number, respectively.
Thus two neighbouring scans result in two independent spectra, 
$T_{ant,j}(\nu,p)$ and $T_{ant,j+1}(\nu,p)$. All the difference
spectra are then averaged, weighted according to their system temperature,
to produce the final difference spectrum for each polarisation.

As an alternative reduction scheme, we used the method of Hainline et al.\ (2006) which involves 
constructing a baseline template for
each {\tt ON} signal from a best-fit linear combination of its neighbouring
{\tt OFF} signals. We can write this as
\begin{equation}
T_{ant,j+1}(\nu,p) = T'_{sys}(\nu,p) \frac{{\tt ON_{i,j+1}}-{\tt OFF^{templ}_{i,j+1}})}{{\tt OFF^{templ}_{i,j+1}})},  \\
\end{equation}
where $T'_{sys}=a\cdot T_{sys,i,j} + b \cdot T_{sys,i,j+2}$ is a combination of the
system temperature of the scans that make up the baseline template:
${\tt OFF^{templ}_{i,j+1}} = a \cdot {\tt OFF_{i,j}}+b\cdot  {\tt OFF_{i,j+2}}$.
The parameters $a$ and $b$ are the best-fit parameters which minimises the quantity
$(({\tt ON_{i,j+1}} - {\tt OFF^{templ}_{i,j+1}})/{\tt OFF^{templ}_{i,j+1}})^2$.
Again, a final spectrum for each polarisation is constructed as a weighted
average of all difference spectra produced in this manner. Vanden Bout et al.\ (2004)
proposed another way of removing problematic baselines, which involves fitting 
a baseline template to the time-averaged spectrum containing the line. However, from a
detailed comparison between the two methods, Hainline et al.\ (2006)
demonstrated that their scheme is just as good at removing baseline-ripples
at the scales one would expect to see a galaxy emission line. As a result we did not attempt to 
apply the method by Vanden Bout et al.\ (2004). 

The final spectra were corrected for atmospheric attenuation using
the opacity values at 23-26\,GHz for the observing dates (R.\ Maddalena, private communication),
and converted from antenna temperature to Janskys by applying an appropriately scaled version of 
the GBT gain-elevation curve at 43\,GHz.

\section{Results \& Discussion}\label{section:results}
\begin{figure*}
\hspace*{-0.3cm}\plotone{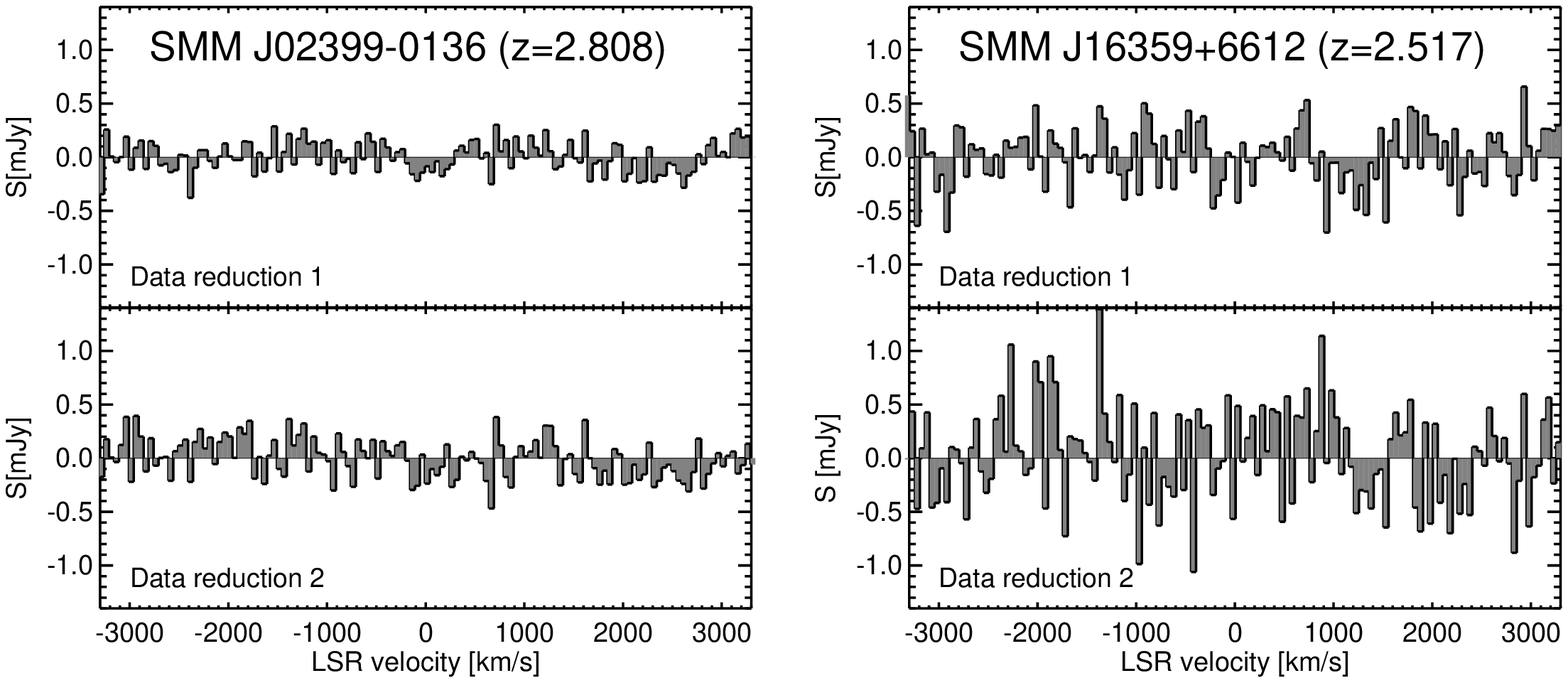}
\caption{HCN$(1-0)$ spectra of J02399 (left) and J16359 (right) 
produced using the GETNOD routine in GBTIDL (top panels) and
our own reduction technique described in \S \ref{section:observations-and-data-reduction}
(bottom panels). The spectra have been plotted on the same flux- and velocity-scale, 
making it easier to compare between the different sources and reduction techniques. 
The LSR velocity-scale is relative to the CO redshift. The spectra have been binned
to $\sim 50$\,km\,s$^{-1}$ and the channel-to-channel noise varies between $0.1-0.5$\,mJy.}
\label{figure:hcn-spectra}
\end{figure*}

The final HCN$(1-0)$ spectra of J02399 and J16359, reduced using the two independent methods
described in the previous section, are shown in Figure \ref{figure:hcn-spectra}. 
The spectra have been smoothed to $50$\,km\,s$^{-1}$ bins. 
Both methods fail to get completely rid of residual baseline wiggles in the final spectra.
However, the spectra reduced with GETNOD appear to be slightly
less noisy than the spectra reduced with the second reduction method. 
The channel-to-channel rms noise of the spectra 
reduced using the GETNOD routine are 0.1 and 0.3\,mJy, respectively.
This is somewhat higher than the theoretical noise
estimates of 0.07 and 0.12\,mJy, calculated for
a typical system temperature $T_{sys}=38\,$K and integration times
corresponding to the ones of J02399 and J16359, respectively.
Thus, it would seem that although the noise does integrate down as $t^{-1/2}$,
the residual baseline wiggles prevents us from reaching the theoretical noise
limit.

No emission is detected significantly above the noise at $V_{LSR}=0$\,km\,s$^{-1}$ (or any
other $V_{LSR}$), which corresponds to the CO redshift and thus the expected 
position of the HCN$(1-0)$ line. However, the sensitivity of the observations allow
us to put upper limits on the HCN$(1-0)$ line luminosity
of these two SMGs.
In doing so we use the upper line flux limits derived from the spectra produced
by GETNOD, as it 
results in the lowest noise spectra. The 3-$\sigma$ upper limits are calculated following
Seaquist, Ivison \& Hall (1995)
\begin{equation}
S_{\mbox{\tiny{HCN(1-0)}}}\Delta V \le 3 \sigma (\delta v \Delta v_{\mbox{\tiny{fwhm}}} ) ^{1/2},
\label{equation:upper-limits}
\end{equation}
where $\sigma$ is the channel-to-channel rms noise, $\delta v$ the velocity
resolution and $\Delta v_{\mbox{\tiny{fwhm}}}$ the line width.
The spectra were binned to a velocity resolution of
50\,km\,s$^{-1}$. We set the HCN$(1-0)$ line widths equal to that of the
CO lines -- see Greve et al.\ (2005). This is probably a conservative estimate since
in local ULIRGs, the HCN line widths are rarely larger than those of CO (Solomon et al.\ 1992; Gao \& Solomon 2004a).
The resulting upper line flux limits are 0.08 and 0.13\,Jy\,km\,s$^{-1}$ for J02399 and J16359,
respectively. From these flux limits we derive upper limits on the apparent HCN$(1-0)$ line luminosities
of $L'_{\mbox{\tiny{HCN(1-0)}}}\le 5.0\times 10^{10}$ and $\le 6.6\times 10^{10}$\,K\,km\,s$^{-1}$pc$^2$ 
for J02399 and J16359, respectively. Correcting for gravitational lensing we find intrinsic
line luminosities of $L'_{\mbox{\tiny{HCN(1-0)}}}\le 2.0\times 10^{10}$ and
$\le 0.3 \times 10^{10}$\,K\,km\,s$^{-1}$pc$^2$.
These upper limits on the HCN line luminosity are similar to those
achieved towards high-$z$ QSOs using the Very Large Array 
(e.g.\ Isaak et al.\ 2004; Carilli et al.\ 2005). 
Table \ref{table:results} lists our findings along with all high-$z$ HCN observations
published in the literature at the time of writing.
It is seen that of 10 sources observed to date only 4 have been 
detected -- the remainder being upper limits. 

In order to determine the star formation efficiency per dense gas mass, and the dense gas
fraction, as gauged by the $L_{\mbox{\tiny{FIR}}}/L'_{\mbox{\tiny{HCN}}}$ and $L'_{\mbox{\tiny{HCN}}}/L'_{\mbox{\tiny{CO}}}$ 
ratio, respectively (Gao \& Solomon 2004a,b), we need accurate estimates of the FIR and CO(1-0)
luminosities.\\
\indent The FIR luminosities of SMGs are generally difficult to estimate accurately,
largely due to the poor sampling of their FIR/submm/radio spectral energy distributions.
Furthermore, in our case the situation is complicated by the fact that most, if not all,
of the sources in Table \ref{table:results} contain AGN, which may be at least partly 
responsible for heating the dust and thus powering the FIR emission. Unfortunately, FIR/submm 
data from the literature do not allow for a detailed modelling of a hot ($\gs 100$\,K) 
dust component (heated by the AGN), and we are therefore unable to correct 
for any AGN contamination. The one exception is the $z=3.9$ BAL quasar,
APM\,08279$+$5255, where the separate AGN vs.\ starburst contributions to the FIR luminosity have
been determined by Rowan-Robinson (2000) who find an apparent FIR luminosity 
of $L_{\mbox{\tiny{FIR}}}\simeq 1.0\times 10^{14}\,\Lsolar$ for the starburst (corrected to the cosmology adopted here). 
In the case of SMGs, extremely deep X-ray observations 
strongly suggest that while every SMG probably harbours an AGN, it is the
starburst which in almost all cases powers the bulk (70-90 percent) of the FIR luminosity (Alexander et al.\ 2005),
and any AGN contamination would therefore not dramatically affect our conclusions.
While optical spectroscopy of J16359 shows no evidence to 
suggest the presence of strong nuclear activity in this source (Kneib et al.\ 2004), this is not the case
in J02399, which appears to be a type-2 QSO judging from its X-ray emission and
optical spectrum properties (Ivison et al.\ 1998; Vernet \& Cimatti 2001). Thus, J02399 could
be a rare case where the AGN contribution is substantial. However, given our inability to correct its FIR
luminosity for AGN contamination, we estimate the FIR luminosities of both sources simply 
by fitting a modified black body with a fixed
$\beta =1.5$ to their rest-frame SEDs and integrating it over the wavelength range $40-120\mu$m, see Table \ref{table:results}. 
This is consistent with the way in which the FIR-luminosities of the objects observed
by Carilli et al.\ (2005) were calculated. The FIR luminosities for the remaining objects in 
Table \ref{table:results} were taken from Carilli et al.\ (2005) and converted to the cosmology
adopted in this paper.\\
\indent The CO luminosities were taken from the compilation of high-$z$ CO detections in 
Greve et al.\ (2005). These are mostly
high-$J$ CO detections ($J=3-2$ or $4-3$), and in order to derive the CO$(1-0)$ line 
luminosity, we assume optically thick, thermalised CO line ratios, i.e.\ 
$(3-2)/(1-0)$ and $(4-3)/(3-2) \sim 1$, see Table \ref{table:results}. It should be noted, however, that studies 
on the ISM in local starburst nuclei yield CO $(3-2)/(1-0) \sim 0.64$ (Devereux et al.\ 1994),
while first results on ULIRGs yield a CO $(6-5)/(4-3)\sim0.6$ ratio
but with CO transitions of $J=4-3$ and higher tracing a
different H$_2$ gas phase than the lower three (Papadopoulos,
Isaak \& van der Werf 2006). These results
along with known examples of  very low global (high$-J$)/(low$-J$) 
CO ratios in high-$z$ starbursts (e.g.\ Papadopoulos \& Ivison 2002; Greve et al.\ 2003; Hainline et al. 2006) suggest
that we may underestimate the SMG CO$(1-0)$ luminosities
when we assume line ratios of unity, especially if
observed CO $J+1\longrightarrow J$, $J+1\ge 4$ lines are used.\\
\indent In Figure \ref{figure:lhcn-lfir} a) -- c) we plot $L_{\mbox{\tiny{FIR}}}$ against
$L'_{\mbox{\tiny{HCN}}}$, $L'_{\mbox{\tiny{HCN}}}$ against $L'_{\mbox{\tiny{CO}}}$, and
$L_{\mbox{\tiny{FIR}}}$ against $L'_{\mbox{\tiny{HCN}}}/L'_{\mbox{\tiny{CO}}}$, respectively,  
for our two sources along with all other existing
high-$z$ HCN observations to date. For comparison we have also plotted the 
sample of local (U)LIRGs (open circles) observed by Gao \& Solomon (2004a,b),
which defines the local $L_{\mbox{\tiny{IR}}}$-$L'_{\mbox{\tiny{HCN}}}$ relation: 
$L_{\mbox{\tiny{IR}}}/L'_{\mbox{\tiny{HCN}}} = 900\,\Lsolar\,$(K\,km\,s$^{-1}$\,pc$^2$)$^{-1}$. 
Since we are interested in the correlations with FIR-luminosity, we carefully fitted modified
black-body SEDs to the Gao \& Solomon sample and derived FIR-luminosities by integrating the
SEDs from $40-120\mu$m. To this end we used all available FIR/submm data in the literature for each
source. The resulting local FIR-HCN correlation is given by 
$\log L_{\mbox{\tiny{FIR}}} = (0.97\pm 0.07) \log L'_{\mbox{\tiny{HCN}}} + (2.9\pm 0.5)$,
and is shown as the solid line in Figure \ref{figure:lhcn-lfir} a). The slope of this correlation is
consistent with unity, suggesting that the local FIR-HCN correlation is linear (Carilli et al.\ 2005).
In the following we shall use this correlation, along with its 1-$\sigma$ limits
allowed by the fitting errors (dotted lines in Figure \ref{figure:lhcn-lfir} a)), to discuss
whether the SMGs and other high-$z$ sources follow the local FIR-HCN correlation.
Secondly, motivated by what is seen in local (U)LIRGs, we shall impose a minimum dense gas fraction 
on the high-$z$ population and from that derive a lower limit on their HCN-luminosity.
Finally, we assume the SMGs follow the local FIR-HCN correlation (within 1-$\sigma$) and
explore what lower limits can be put on their dense gas fraction.

With only four HCN detections to date and little more than a handful of upper limits,
including the ones presented here, it is difficult to determine whether the local
$L_{\mbox{\tiny{FIR}}} - L'_{\mbox{\tiny{HCN}}}$ relation extends to higher redshifts and luminosities. 
It is interesting to note in Figure \ref{figure:lhcn-lfir} a), however, 
that while the three SMGs with HCN observations (our two sources plus SMM\,J14011$+$0252)
all are consistent within $1$-$\sigma$ of the local relation extrapolated to higher luminosities, 
5/7 of the QSOs observed in HCN fall above the $1$-$\sigma$ envelope.
This may suggest that QSOs have higher FIR luminosities for a fixed HCN luminosity than the SMGs. 
Although, this is not statistically significant,
it is consistent with the expected picture in which the AGN in QSOs contribute
a higher fraction to the FIR luminosity than in SMGs. Correcting for the AGN contribution
in QSOs would lower them from where they are currently plotted, bringing them more
in line with the local $L_{\mbox{\tiny{FIR}}} - L'_{\mbox{\tiny{HCN}}}$ correlation.\\
\indent Alternatively, the data
could in principle also be interpreted as a steepening of the slope of the correlation at higher luminosities,
meaning that the AGN contribution in both QSOs and SMGs is higher than in 
local (U)LIRGs -- a scenario which would be in line with the suggested increase in 
AGN dominance with FIR luminosity in local ULIRGs (Veilleux et al.\ 1999; Tran et al.\ 2001).
Although, X-ray constraints suggest that AGN contribute $\ls 30$ percent of to the FIR
luminosity of SMGs, they do not allows us to completely rule out the above scenario since
it is possible many SMGs harbour obscured, Compton-thick AGN (Alexander et al.\ 2005). 
However, the fact that the star formation
rates for SMGs derived from reddening-corrected H$\alpha$ luminosities agree with those
derive from their FIR luminosities, suggest that AGN are unlikely to contribute signficantly to the
FIR.

We see from Figure \ref{figure:lhcn-lfir} b) that all the high-$z$ objects
lie above the $L'_{\mbox{\tiny{HCN}}} - L'_{\mbox{\tiny{CO}}}$ correlation fitted
to local galaxies with FIR luminosities less than $10^{11}\,\Lsolar$ and with a fixed slope of unity. 
This is consistent with the the steepening of the $L'_{\mbox{\tiny{HCN}}} - L'_{\mbox{\tiny{CO}}}$ 
correlation at higher FIR luminosities seen within the local (U)LIRGs sample itself
as reported by Gao \& Solomon (2004b). They argued that the dense molecular gas fraction,
as gauged by the $L'_{\mbox{\tiny{HCN}}}/L'_{\mbox{\tiny{CO}}}$ ratio, is a powerful
indicator of starburst dominated systems. Also, locally they found that 
all galaxies with $L'_{\mbox{\tiny{HCN}}}/L'_{\mbox{\tiny{CO}}} \ge 0.06$ 
have $L_{\mbox{\tiny{IR}}} \gs 10^{11}\,\Lsolar$. In Figure \ref{figure:lhcn-lfir} c) we
have plotted our estimates of $L_{\mbox{\tiny{FIR}}}$ for the local (U)LIRG sample against their
$L'_{\mbox{\tiny{HCN}}}/L'_{\mbox{\tiny{CO}}}$ ratios, and find that apart from two sources,
the same criterion can be applied to the FIR luminosity, i.e.\ sources with $L'_{\mbox{\tiny{HCN}}}/L'_{\mbox{\tiny{CO}}} \ge 0.06$ 
have $L_{\mbox{\tiny{FIR}}} \gs 10^{11}\,\Lsolar$. 
It is also seen that all the high-$z$ sources observed so far -- all of which have been very luminous -- obey the same criterium.
Given their large FIR luminosities, it therefore seems reasonable to assume that the SMGs too
have high dense gas fractions, i.e.\ $L'_{\mbox{\tiny{HCN}}}/L'_{\mbox{\tiny{CO}}}\ge 0.06$. 
In that case, one can put a {\it lower} limit on their
HCN luminosity and investigate where they would lie in the FIR-HCN diagram.
The range of allowed HCN luminosities derived in this way for each SMG (and other high-$z$ objects) is shown as 
horizontal lines in Figure \ref{figure:lhcn-lfir}a. It is seen that while J02399 and J16359, and the other high-$z$ objects,
could lie above the upper dotted line and still have a high dense gas fraction ($L'_{\mbox{\tiny{HCN}}}/L'_{\mbox{\tiny{CO}}}\ge 0.06$), 
this is not the case for J14011. Thus, if our above assumption is correct, we can conclude that at least one of the SMGs (J14011), 
and possibly more, is consistent with the local FIR-HCN correlation, and that this SMG should be detectable in HCN if
the sensitivity is improved by a factor $\times 2$.

APM\,08279$+$5255 has the highest measured dense gas fraction, based on detections, in the high-$z$ sample, even
though it falls above the local $L_{\mbox{\tiny{FIR}}} - L'_{\mbox{\tiny{HCN}}}$ relation in
Figure \ref{figure:lhcn-lfir} a). Naively, this would indicate that the ISM in this system
is dominated by dense gas, and that star formation contributes substantially to its
FIR-luminosity. 
However, caution should be taken since we have
calculated the HCN/CO ratio using the nuclear CO line luminosity of APM\,08279
(Papadopoulos et al.\ 2001). Adopting the global CO line luminosity, which includes emission extended
on 10-kpc scales (Papadopoulos et al.\ 2001), i.e.\ scales comparable to those probed by the HCN observations 
({\sc fwhm}~$\simeq 7\arcsecs$  -- Wagg et al.\ 2005), results in a HCN/CO ratio almost an order
of magnitude lower. 
Similarly, we see that the upper limits on
the dense molecular gas fraction in J02399 and J16359 are rather high, yet both
sources lie close to the $L_{\mbox{\tiny{FIR}}} - L'_{\mbox{\tiny{HCN}}}$ correlation
in Figure \ref{figure:lhcn-lfir} a). What would the dense gas fraction of the SMGs be if they were required to lie between
the upper $1$-$\sigma$ line of the local FIR-HCN correlation and their observed HCN limits (for
unchanged FIR luminosities)? The answer is given in 
Figure \ref{figure:lhcn-lfir} c), where the permitted range in HCN/CO luminosity ratios are
marked as horizontal lines.
It is seen that J14011 has a dense gas fraction well within the range of most ULIRGs, and could
have $L'_{\mbox{\tiny{HCN}}}/L'_{\mbox{\tiny{CO}}} \ls 0.06$ and still be consistent with the FIR-HCN
correlation. For the other two SMGs we see that even in the most conservative case, where the SMGs are
just barely consistent with the $1$-$\sigma$ line, they would still have
some of the highest dense molecular gas fractions observed, comparable to 
the most extreme starburst dominated ULIRGs seen locally. Thus we conclude that if the SMGs
do in fact follow the FIR-HCN correlation (or are within $1$-$\sigma$ of it), then J02399 and J16359
are likely to have higher dense gas fractions than J14011.
Note, however, that if an AGN contributes significantly
to the FIR luminosity, the above argument would yield a higher HCN luminosity, and thus higher
HCN/CO ratios than its true value, but we emphasize that the energy output at FIR/submm wavelengths 
from AGN in SMGs is found to be negligble compared to the starburst (Alexander et al.\ 2005).

\section{Summary}
\begin{figure}
\plotone{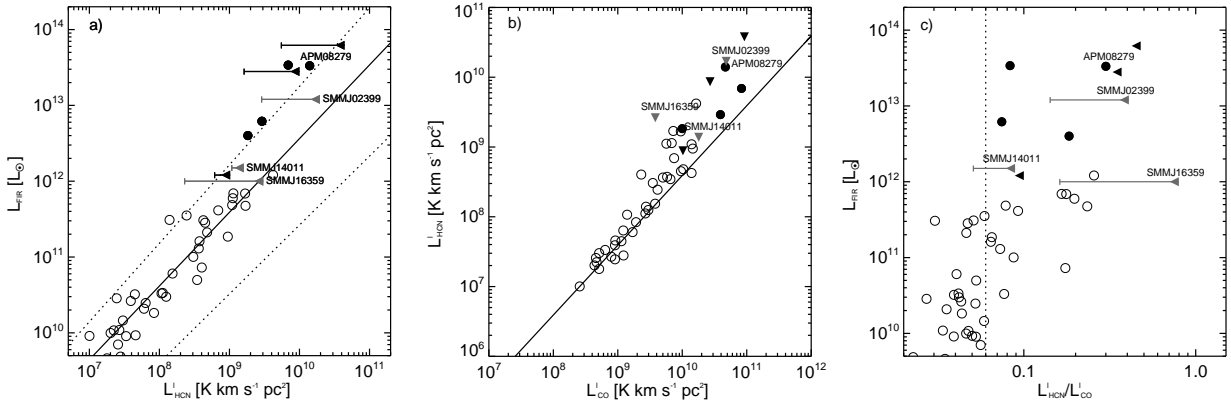}
\caption{Correlations between different combinations of FIR, HCN and CO luminosity for local (U)LIRGs (open circles - 
Gao \& Solomon 2004a,b) and the 10 high-$z$ objects observed in HCN to date (solid
circles and triangles). The latter have been corrected for gravitational lensing where appropriate, see Table \ref{table:results}.
The three grey triangles denote upper limits of the two SMGs observed as part of this study and
SMM\,J14011$+$0210 which was oberved in HCN by Carilli et al.\ (2005) using the Very Large Array. 
The solid line in a) represents the local FIR-HCN correlation derived in this paper:
$\log L_{\mbox{\tiny{FIR}}}=  0.97 \log L'_{\mbox{\tiny{HCN}}} + 2.8$, while the horizontal
extensions to the upper limits mark the $L'_{\mbox{\tiny{HCN}}}$-range allowed for each source given its
observed upper HCN limit and assuming that $L'_{\mbox{\tiny{HCN}}}/L'_{\mbox{\tiny{CO}}} \ge 0.06$.
The solid line in b) is a fit (with a fixed slope of 1) to normal spiral galaxies 
in the local Universe. The dotted line in c) marks the 
$L'_{\mbox{\tiny{HCN}}}/L'_{\mbox{\tiny{CO}}}$ limit above which sources
in the Gao \& Solomon (2004a,b) sample are (U)LIRGs. The range in HCN/CO luminosity
ratios for the SMGs, if they were to lie between the upper $1$-$\sigma$ line of the local 
$L_{\mbox{\tiny{FIR}}}-L'_{\mbox{\tiny{HCN}}}$ correlation, and their observed 
HCN limits are shown as grey horizontal lines.}
\label{figure:lhcn-lfir}
\end{figure}

With this study we have increased the number of submm-selected galaxies with 
HCN$(1-0)$ observations from 1 to 3, although the upper limit on the HCN luminosity
of one of our sources (J16359) is not a very sensitive one.
The upper limits on the HCN$(1-0)$ line luminosity of these three SMGs 
are consistent with the $L_{\mbox{\tiny{FIR}}}-L'_{\mbox{\tiny{HCN}}}$ relation observed in
the local Universe (Gao \& Solomon 2004a,b). In order to test whether a typical SMG with
a FIR-luminosity of $L_{\mbox{\tiny{FIR}}}\sim 10^{12.5}\,\Lsolar$ is inconsistent with the local relation,
we need to be sensitive to HCN luminosities of $L'_{\mbox{\tiny{HCN}}} \ls 10^{10}\,$K\,km\,s$^{-1}$\,pc$^2$.
Thus, we appear to be close (less than a factor of two) in sensitivity to that
needed from Figure \ref{figure:lhcn-lfir} in order to detect HCN in SMGs. 
We conclude that until deeper observations come along there is no evidence
to suggest that SMGs have higher star formation efficiencies per unit dense gas 
mass than local (U)LIRGs or indeed Galactic GMCs (Wu et al.\ 2005).


\acknowledgments
We thank the referee for useful comments which helped improve the paper
substantially. We are grateful to NRAO for financial and scientific support, and in particular
to the telescope operators at Green Bank for their expertise. We thank
Ron Maddalena and Toney Minter for their help in reducing the data and
for providing the opacity values used. LJH was supported 
by the GBT graduate student funding program. AWB acknowledges support
from the Alfred P.\ Sloan Foundation and the Research Corporation. IRS acknowledges 
support from the Royal Society.

\clearpage

\clearpage

\begin{deluxetable}{llllllllll}
\rotate
\tabletypesize{\scriptsize}
\tablecaption{Observations of HCN at high redshifts to date. The two SMGs studied here are in the upper part of the table. The luminosities have not been corrected for gravitational
lensing, but the lensing amplifications ($\mu$) are given in column 9.}
\tablewidth{0pt}
\tablehead{
\colhead{Source}      & \colhead{Type} & \colhead{$z$}  & \colhead{HCN line} & \colhead{$L'_{\mbox{\tiny{HCN}}}$}                & \colhead{CO line} & \colhead{$L'_{\mbox{\tiny{CO}}}$}              & \colhead{$L_{\mbox{\tiny{FIR}}}$}    & \colhead{$\mu$} & \colhead{Ref.} \\
\colhead{}            & \colhead{}     & \colhead{}     & \colhead{}         & \colhead{$\times 10^{10}\,$K\,km\,s$^{-1}$pc$^2$} & \colhead{}        & \colhead{$\times 10^{10}\,$K\,km\,s$^{-1}$pc$^2$} & \colhead{$\times 10^{12}\,\Lsolar$} & \colhead{}      & \colhead{}     
}
\startdata
SMM\,J02399$-$0136    & SMG            & 2.808          & $1-0$              & $\le 4.9 $                                       &$3-2$              & $12\pm 2$                                         & $30 $                               & $2.5$           & [1],[2],[3]\\
SMM\,J16359$+$6612    & SMG            & 2.517          & $1-0$              & $\le 6.7 $                                       &$3-2$              & $8.4\pm 0.4$                                      & $22$                                & $22$            & [1],[4],[5]\\
\tableline
SMM\,J14011$+$0252    & SMG            & 2.565          & $1-0$              & $\le 0.8$                                        &$3-2$              & $9.0\pm 0.7$                                      & $8$                               & $5$             & [6],[7],[8]\\
SDSS\,J1148$+$5251    & QSO            & 6.419          & $1-0$              & $\le 1.0$                                        &$3-2$              & $2.7\pm 0.6$                                      & $28$                                & $1$             & [6],[9],[10]\\
BR\,1202$-$0725       & QSO            & 4.693          & $1-0$              & $\le 4.4$                                        &$2-1$              & $9.2 \pm 1.0$                                     & $62$                                & $1$             & [11],[12]\\
APM\,08279$+$5255     & QSO            & 3.911          & $5-4$              & $4.2\pm 0.5$                                     &$1-0$              & $14\pm 3$                                         & $100$                               & $3^a$           & [13],[14],[15]\\       

MG\,0751$+$2716       & QSO            & 3.200          & $1-0$              & $\le 1.7$                                        &$4-3$              & $17\pm 1$                                     & $20$                                & $16.6$          & [6],[16]\\
QSO\,J1409$+$5628     & QSO            & 2.583          & $1-0$              & $0.7\pm 0.2$                                   &$3-2$              & $8.3\pm 0.3$                                      & $34$                                & $1$             & [6],[17],[18]\\
H1413$+$117           & QSO            & 2.558          & $1-0$              & $3.2\pm 0.6$                                     &$3-2$              & $43\pm 7$                                     & $68$                                & $11$            & [19],[18],[19],[20],[21]\\
                      & . . .          & . . .          & $4-3$              & $8.6\pm 1.6$                                     &                   & . . .                                             & . . .                               & . . .           & [22]\\
IRAS\,F10214$+$4724   & QSO            & 2.286          & $1-0$              & $2.2\pm 0.4$                                     &$3-2$              & $12\pm 3$                                     & $48$                                & $13$            & [23],[24]\\
\enddata
\tablerefs{
       [1] This work; [2] Frayer et al.\ (1998); [3] Genzel et al.\ (2003); [4] Sheth et al.\ (2004); [5] Kneib et al.\ (2004);
       [6] Carilli et al.\ (2005); [7] Frayer et al.\ (1999); [8] Downes \& Solomon (2003); [9] Bertoldi et al.\ (2003); [10] Walter et al.\ (2003);
       [11] Isaak et al.\ (2004); [12] Carilli et al.\ (2002); [13] Wagg et al.\ (2005); [14] Papadopoulos et al.\ (2001); 
       [15] Lewis et al.\ (2002); [16] Barvainis et al.\ (2002); [17] Beelen et al.\ (2004); [18] Hainline et al.\ (2004); [19] Solomon et al.\ (2003); [20] Wei\ss\ et al.\ (2003); 
       [21] Wilner et al.\ (1995); [22] Barvainis et al.\ (1997); [23] Vanden Bout et al.\ (2004); [24] Solomon et al.\ (1992).
}
\label{table:results}
\tablenotetext{a}{We adopt the lensing amplification derived from high-resolution CO maps by Lewis et al.\ (2002). Note, however, the
amplification factor may be as high as $\mu \sim 7-20$ (Downes et al.\ 1999)
}
\end{deluxetable}







\begin{thebibliography}{}
    \bibitem[Alexander et al. (2005)]{Alexander-et-al-2005} 
      Alexander, D.\ M., Bauer, F.\ E., Chapman, S.\ C., Smail, I., Blain, A.\ W., Brandt, W.\ N., Ivison, R.\ J.\ 2005, ApJ, 632, 736.
    \bibitem[Barvainis et al. (1997)]{Barvainis-et-al-1997} 
      Barvainis, R., Maloney, P., Antonucci, R., Alloin, D.\ 1997, ApJ, 484, 695.
    \bibitem[Barvainis et al. (2002)]{Barvainis-et-al-2002} 
      Barvainis, R.,  Alloin D., Bremer, M.\ 2002, A\&A, 385, 399.
    \bibitem[Beelen et al. (2004)]{Beelen-et-al-2004} 
      Beelen, A.\ et al.\ 2004, A\&A,423, 441.
    \bibitem[Bertoldi et al. (2003)]{Bertoldi-et-al-2003} 
      Bertoldi, F., et al.\ 2003, A\&A, 409, L47.
    \bibitem[Blain et al. (2002)]{Blain-et-al-2002} 
      Blain, A.\ W., et al.\ 2002, PhR, 369, 111.
    \bibitem[Carilli et al. (2002)]{Carilli-et-al-2002} 
      Carilli, C.\ L., et al.\ 2002, ApJ, 575, 145.
    \bibitem[Carilli et al.(2005)]{Carilli-et-al-2005} 
      Carilli, C.\ L., et al.\ 2005, ApJ, 618, 586.
    \bibitem[Devereux et al. (1994)]{Devereux-et-al-1994} 
      Devereux, N., Taniguchi, Y., Sanders, D.\ B., Nakai, N., Young, J.\ S.\ 1994, AJ, 107, 2006.
    \bibitem[De Breuck et al. (2005)]{De-Breuck-et-al-2005} 
      De Breuck, C., Downes, D., Neri, R., van Breugel, W., Reuland, M., Omont, A., Ivison, R.\ 2005, A\&A, 430, L1.
    \bibitem[Downes et al. (1999)]{Downes-et-al-1999} 
      Downes, D., Neri, R., Wiklind, T., Wilner, D.\ J., Shaver, P.\ A.\ 1999, ApJ, 513, L1.
    \bibitem[Frayer et al. (1998)]{Frayer-et-al-1998} 
      Frayer, D.\ T., Ivison, R.\ J., Scoville, N.\ Z., Yun, M., Evans, A.\ S., Smail, I., Blain, A.\ W., Kneib, J.-P.\ 1998, ApJ, 506, L7.
    \bibitem[Frayer et al. (1999)]{Frayer-et-al-1999} 
      Frayer, D.\ T., et al.\ 1999, ApJ, 514, L13.
    \bibitem[Gao \& Solomon (2004a)]{Gao-and-Solomon-2004a} 
      Gao, Y.\ \& Solomon, P.\ M.\ 2004a, ApJS, 152, 63.
    \bibitem[Gao \& Solomon (2004b)]{Gao-and-Solomon-2004b} 
      Gao, Y.\ \& Solomon, P.\ M.\ 2004b, ApJ, 606, 271. 
    \bibitem[Genzel et al. (2003)]{Genzel-et-al-2003} 
      Genzel, R., Baker, A.\ J., Tacconi, L.\ J., Lutz, D., Cox, P., Guilloteau, S., Omont, A.\ 2003. ApJ, 584, 633.
    \bibitem[Greve et al. (2003)]{Greve-et-al-2003} 
      Greve, T.\ R., Ivison, R.\ J., Papadopoulos, P.\ P.\ 2003, ApJ, 599, 839.
    \bibitem[Greve et al. (2005)]{Greve-et-al-2005} 
      Greve, T.\ R., et al.\ 2005, MNRAS, 359, 1165.
    \bibitem[Hainline et al. (2004)]{Hainline-et-al-2004} 
      Hainline, L.\ J., Scoville, N.\ Z., Yun, M.\ S., Hawkins, D.\ W., Frayer. D.\ T., Isaak, K.\ G., 2004, ApJ, 609, 61.
    \bibitem[Hainline et al. (2006)]{Hainline-et-al-2006} 
      Hainline, L.\ J., et al.\ 2006, ApJ, submitted.
    \bibitem[Isaak et al. (2004)]{Isaak-et-al-2004} 
      Isaak, K.\ G., Chandler, C.\ J., Carilli, C.\ L.\ 2004, MNRAS, 348, 1035.
    \bibitem[Ivison et al. (1998)]{Ivison-et-al-1998} 
	Ivison, R.\ J., Smail, Ian, Le Borgne, J.-F., Blain, A.\ W., Kneib, J.-P., Bezecourt, J., Kerr, T.\ H., Davies, J.\ K.\ 1998, MNRAS, 298, 583.
    \bibitem[Kneib et al. (2004)]{Kneib-et-al-2004} 
      Kneib, J.-P., van der Werf, P., Knudsen, K.\ K., Smail, Ian, Blain, A.\ W., Frayer, D., Barnard, V., Ivison, R.\ 2004, MNRAS, 349, 1211.
    \bibitem[Kneib et al.\ (2005)]{Kneib-et-al-2005} 
      Kneib, J.-P., Neri, R., Smail, I., Blain, A.\ W., Sheth, K., van der Werf, P., Knudsen, K.\ K.\ 2005, A\&A, 434, 819.
    \bibitem[Neri et al. (2003)]{Neri-et-al-2003} 
      Neri, R., et al.\ 2003, ApJ, 597, L113.
    \bibitem[Papadopoulos et al. (2000)]{Papadopoulos-et-al-2000}  
      Papadopoulos, P.\ P., R\"ottgering, H.\ J.\ A., van der Werf, P.\ P., Guilloteau, S., Omont, A., van Breugel, W.\ J.\ M. \& Tilanus, R.\ P.\ J.\ 2000, ApJ, 528, 626.
    \bibitem[Papadopouloss et al. (2001)]{Papadopoulos-et-al-2001} 
      Papadopoulos, P.\ P., Ivison, R.\ J., Carilli, C.\ L., Lewis, G.\ 2001, Nature, 409, 58. 
    \bibitem[Papadopouloss \& Ivison (2002)]{Papadopoulos-and-Ivison-2002} 
      Papadopoulos, P.\ P.\ \& Ivison, R.\ J.\ 2002, ApJ, 564, L9.
    \bibitem[Papadopoulos, Isaak \& van der Werf (2006)]{Papadopoulos-et-al-2006} 
      Papadopoulos, P.\ P., Isaak, K.\ G.\ \& van der Werf, P.\ 2006, ApJ, submitted.
    \bibitem[Rowan-Robinson et al. (2000)]{Rowan-Robinson-et-al-2000} 
      Rowan-Robinson, M.\ 2000, MNRAS, 316, 885.
    \bibitem[Seaquist et al. (1995)]{Seaquist-et-al-1995} 
      Seaquist, E.\ R., Ivison, R.\ J., Hall, P.\ J.\ 1995, MNRAS, 276, 867.
    \bibitem[Sheth et al. (2004)]{Sheth-et-al-2004} 
      Sheth, K., Blain, A.\ W., Kneib J.-P., Frayer, D.\ T., van der Werf, P., Knudsen, K.\ K.\ 2004, ApJ, 614, L5.
    \bibitem[Smail et al.\ (1997)]{Smail-et-al-1997} 
      Smail, I., Ivison, R.\ J., Blain, A.\ W.\ 1997, ApJ, 490, L5.
    \bibitem[Smail et al. (2002)]{Smail-et-al-2002} 
      Smail, I., Ivison, R.\ J., Blain, A.\ W., Kneib, J.-P.\ 2002, MNRAS, 331, 495.
    \bibitem[Solomon et al. (1992)]{Solomon-et-al-1992} 
      Solomon, P.\ M., Downes, D., Radford, S.\ J.\ E.\ 1992, ApJ, 387, L55.
    \bibitem[Solomon et al. (2003)]{Solomon-et-al-2003} 
      Solomon, P.\ M., Vanden Bout, P., Carilli, C.\ L., Guelin, M., 2003, Nature, 426, 636.
    \bibitem[Solomon \& Vanden Bout (2005)]{Solomon-and-Vanden-Bout-2005} 
      Solomon, P.\ \& Vanden Bout, P.\ 2005, ARAA, 43, 677.
    \bibitem[Spergel et al. (2003)]{Spergel-et-al-2003} 
      Spergel, D.\ N., et al.\ 2003, ApJS, 148, 175.
    \bibitem[Tacconi et al. (2006)]{Tacconi-et-al-2006} 
      Tacconi, L., et al.\ 2006, ApJ, 640, 228.
    \bibitem[Tran et al. (2001)]{Tran-et-al-2001} 
      Tran, Q.\ D.\ et al.\ 2001, ApJ, 552, 527.
    \bibitem[Vanden Bout et al. (2004)]{Vanden-Bout-et-al-2004} 
      Vanden Bout, P.\ A., Solomon, P.\ M., Maddalena, R.\ J.\ 2004, ApJ, 614, L97. 
    \bibitem[Veilleux et al. (1999)]{Veilleux-et-al-1999} 
      Veilleux, S., Kim, D.-C., Sanders, D.\ B.\ 1999, ApJ, 522, 113.
    \bibitem[Vernet \& Cimatti (2001)]{Vernet-and-Cimatti-2001} 
      Vernet, J.\ \& Cimatti, A.\ 2001, A\&A, 380, 409.
    \bibitem[Wagg et al. (2005)]{Wagg-et-al-2005} 
      Wagg, J., Wilner, D.\ J., Neri, R., Downes, D., Wiklind, T.\ 2005, ApJ, 634, L13.
    \bibitem[Walter et al. (20030]{Walter-et-al-2003} 
      Walter, F., et al.\ 2003, Nature, 424, 406.
    \bibitem[Weiss et al. (2005)]{Weiss-et-al-2005} 
      Wei\ss, A., Downes, D.,  Henkel, C.\ 2005, A\&A, 440, L45.
    \bibitem[Wilner et al. (1995)]{Wilner-et-al-1995} 
      Wilner, D.\ J., Zhao, J.-H., Ho, P.\ T.\ P., 1995, ApJ, 453, L91.
    \bibitem[Wu et al. (2005)]{Wu-et-al-2005} 
      Wu, J., Evans, N.\ J., Gao, Y., Solomon, P.\ M., Shirley, Y.\ L., Vanden Bout, P.\ A.\ 2005, ApJ, 635, L173.
\end{thebibliography}
\end{document}